# Cross-vehicle injury spillovers of ADAS functions in two-vehicle rear-end and same-direction sideswipe crashes

Parvez Anowar[a,*], Mohamed Abdel-Aty[a], Chenzhu Wang[a], Shaoyan Zhai[a]

[a] Department of Civil, Environmental and Construction Engineering, University of Central Florida, Orlando, FL 32816, United States

* Corresponding author.
E-mail address: pa545735@ucf.edu (P. Anowar).

## Abstract

Advanced driver assistance system (ADAS) functions are evaluated mainly by the crashes they prevent, and little is known about how they relate to the injuries sustained in the crashes that would still occur. Because two colliding vehicles share one impact, a function on one vehicle may also be associated with the other driver's injury. This study examines whether an ADAS function on either vehicle in such two-vehicle crashes is associated with either driver's injury. It focuses on rear-end (n = 2,273) and same-direction sideswipe crashes (n = 1,216) from Texas police reports across 2023 to 2025. Striking and struck roles were assigned within each crash, and the ADAS functions of each vehicle were identified from vehicle identification numbers. For each crash type, a bivariate probit model, with random parameters for rear-end crashes, jointly estimates whether each driver was injured. Significant cross-equation correlations (0.559 and 0.375) support the joint specification. Electronic stability control on the striking vehicle is associated with a lower struck-driver injury probability in both crash types, by 0.036 to 0.073 in rear-end and 0.102 in sideswipe crashes. Lane departure warning on the striking vehicle shows a similar association in sideswipe crashes (about 0.078). Traction control with anti-lock braking on the struck vehicle shows the largest association among the ADAS functions with its own driver's injury probability (0.097 and 0.103 lower). No struck-vehicle function is associated with the striking driver's injury, and forward-looking functions show no association with either driver's injury, consistent with their benefit lying mainly in crash prevention. The findings suggest that the protective benefits of stability and lane departure functions may reach beyond the equipped vehicle to the other driver, strengthening the case for their wider fleet adoption.

## 1. Introduction

Traffic crashes take a heavy toll on people and on the economy. In the United States, motor vehicle traffic crashes killed 40,901 people in 2023 and left an estimated 2.44 million people injured (National Center for Statistics and Analysis 2025). The economic cost of the motor vehicle crashes of 2019 was estimated at $340 billion, and their total societal harm, which also counts the value of lost quality of life, was estimated at nearly $1.4 trillion (Blincoe et al. 2023). Among the most common of these events are collisions between two vehicles traveling in the same direction, especially rear-end crashes, which are the most frequent crash type on United States roads (Russo et al. 2023). Same-direction sideswipes are the other principal crash type. Because crashes of both kinds are rarely fatal, their cost comes largely from the injuries they produce, making injury severity a natural focus once a crash has occurred. Understanding which characteristics of the two vehicles and their drivers are associated with those injuries is useful for both vehicle technology policy and traffic safety countermeasures.

Each crash of this kind produces two injury outcomes, one for each driver (apart from passengers), and the two need not be equally severe (Zou et al. 2023). Studies that resolve the striking and the struck driver find that injury severity and its associated factors differ by role (Yuan et al. 2020; Russo et al. 2023; Wang et al. 2024c). One vehicle's body type, mass, and age are also associated with the other driver's crash risk and injury (Abdel-Aty and Abdelwahab 2004a, 2004b; Hoye 2019; Dabbour et al. 2020). Whether a vehicle's crash-avoidance or driver assistance functions reach the other driver in the same way remains largely an open question.

New vehicles now have advanced driver assistance system (ADAS) functions that help drivers avoid crashes, and several of them act on the two crash types considered here. Forward-looking functions, such as forward collision warning and automatic emergency braking, respond to the risk of a rear-end collision with the vehicle ahead. Lateral functions, such as lane departure warning, act on the drift out of the lane that can lead to a sideswipe. Stability functions, such as electronic stability control, help the driver keep control during hard braking or evasive steering (Erke 2008; Sivinski 2011). These functions have been evaluated mostly by whether equipped vehicles are involved in fewer crashes (Erke 2008; Cicchino 2017, 2018a), not by the probability of injury when a crash is not prevented (the severity margin). Injuries in such crashes depend on the speed, maneuver, and angle at which the vehicles meet, and on whether each vehicle stays under control through the contact. A function may influence these conditions even in a crash it does not prevent.

Paired outcomes also shape how severity should be modeled. A joint modeling approach treats the two drivers' outcomes as one system, resolving the pair into an at-fault and a not-at-fault driver and allowing the two severity equations to share unobserved factors (Russo et al. 2014, 2023; Wali et al. 2018). These models, however, have used mainly the usual driver, vehicle, and roadway variables, and none appears to have included individual ADAS functions in the severity equation of either driver.

This study addresses this gap with separate models for police-reported two-vehicle rear-end and same-direction sideswipe crashes in Texas. To the authors' knowledge, no injury-severity model has been reported for the same-direction sideswipe alone, so the sideswipe estimates add role-resolved evidence for an understudied crash type. For each crash type, the striking and struck drivers' injury outcomes are estimated jointly through a bivariate probit model linked by a cross-equation error correlation, with random parameters for rear-end crashes and fixed parameters for sideswipe crashes. The ADAS functions of both vehicles are identified at function level from each vehicle's identification number. The study examines whether the factors associated with injury differ between the striking and struck drivers. It also examines whether a function on one vehicle is associated with the other driver's injury as well as its own driver's, and whether these associations differ between the two crash types. Throughout, the estimates are interpreted as associations, not as effects of the functions.

The contribution is a cross-vehicle, function-level analysis of ADAS availability at the severity margin. A function on the striking vehicle may be associated with its own driver's injury (self-protection) or with the struck driver's injury (spillover). A function on the struck vehicle may be associated with its own driver's injury (the struck driver's self-protection) or with the striking driver's injury (a reverse spillover). The two spillover perspectives concern the other vehicle's driver, so the benefits at stake extend beyond the buyers of the equipped vehicles.

The remainder of the paper first reviews the literature on injury severity by role, ADAS effectiveness, and joint models, and then describes the data and function coding. It next presents the modeling framework, reports the estimates and marginal effects for both crash types, and concludes with implications and limitations.

## 2. Literature review

### *2.1. Injury severity by role and cross-vehicle influence*

A two-vehicle crash produces two injury outcomes at once, yet most severity research has examined one vehicle's outcome at a time. In intersection-related rear-end crashes, not-at-fault drivers had clearly higher shares of possible and evident injury than at-fault drivers, and several factors entered the two drivers' equations differently or entered only one of them (Russo et al. 2023). Wang et al. (2024c) found both shared and role-specific determinants in freeway rear-end crashes, and Dabbour et al. (2020) showed that a driver's own age and sex raised that driver's severity but not the other's.

The same-direction sideswipe has received less attention in severity research than the rear-end crash and is usually combined with the opposite-direction sideswipe in broader samples. Wang et al. (2015) pooled same-direction and opposite-direction sideswipes into one category, and in Chiou et al. (2013) the sideswipe was not a significant predictor of injury severity and its two directions were not distinguished. Because these crash types involve different maneuvers, points of impact, and vehicle technologies, findings from rear-end crashes or pooled sideswipes cannot simply be extended to the same-direction sideswipe (Hua and Fan 2023; Scharber 2024).

Prior studies have long found that one vehicle's characteristics and its driver's behavior are associated with the other driver's injuries. In two-vehicle side impacts, a higher speed limit on the striking vehicle's road was associated with sharply higher serious-injury odds for the struck occupants (Farmer et al. 1997). Light trucks, vans, and sport utility vehicles are associated with lower fatality risk for their own occupants and higher fatality risk for the other vehicle's occupants (Abdel-Aty and Abdelwahab 2004a; Abdelwahab and Abdel-Aty 2004). In rear-end crashes, an older vehicle was associated with higher injury severity for the driver it collided with (Dabbour et al. 2020), and in car-to-car collisions a driver's serious-injury risk was lower when the other car was newer (Hoye 2019). The at-fault driver's fatigue or intoxication has likewise been linked to the other driver's severity (Chiou et al. 2013; Wali et al. 2018).

These findings require a severity model that resolves the striking and struck roles into separate equations and includes both vehicles' attributes in each. Yet the cross-vehicle attributes examined in these severity studies have all been familiar driver, vehicle, or crash characteristics, and none has been a crash-avoidance or driver assistance technology.

### *2.2. ADAS effectiveness and injury severity*

Most evaluations of the ADAS functions measure whether crashes occur, including injury and fatal crashes, rather than how severe the injuries are. Table 1 summarizes these evaluations by function, with each study's data and design, the margin it evaluates, and its principal estimates. Electronic stability control (ESC) has the most extensive record, with a meta-analysis estimating roughly 50% fewer single-vehicle crashes (Erke 2008) and other studies from the United States and Sweden pointing the same way (Farmer 2006; Lie et al. 2006; Sivinski 2011). Its benefit is less certain for non-fatal multi-vehicle crashes, where the estimates are smaller and mostly non-significant (Hoye 2011). Forward collision warning and automatic emergency braking have the strongest record for preventing rear-end crashes, with front-to-rear crash involvement 27% lower with warning alone and 50% lower with automatic braking added (Cicchino 2017). Simulation suggests that these systems would prevent few rear-end crashes, but the prevented crashes and the lower impact speeds from braking before impact together would leave 29% to 50% fewer belted drivers with moderate or more severe injuries (Kusano and Gabler 2012).

**Table 1 Summary of evaluations of the ADAS functions**

| Function group | Study | Data and design | Margin evaluated | Principal estimates |
|---|---|---|---|---|
| Stability (ESC) | Farmer (2004) | United States, same-model vehicles with and without ESC | Crash occurrence | Single-vehicle -41% (CI -48, -33); fatal single-vehicle -56%; overall -7% |
| Stability (ESC) | Farmer (2006) | United States, update; cars and SUVs | Crash occurrence | Single-vehicle -41% (SUV -49%, car -33%); multi-vehicle fatal -37% (SUV) and -25% (car, short of significance) |
| Stability (ESC) | Lie et al. (2006) | Sweden, induced exposure; injury crashes excluding rear-end | Injury-crash occurrence | All injury crashes -16.7% (95% CI half-width 9.3); serious and fatal -21.6% (half-width 12.8); wet-road serious and fatal loss-of-control -56.2% |
| Stability (ESC) | Erke (2008) | Meta-analysis of eight effectiveness studies; single-vehicle, head-on, and multi-vehicle configurations | Crash occurrence | Single-vehicle -49% (CI -55, -42); multi-vehicle fatal -32% (CI -43, -20); head-on -13% (CI -17, -8) |
| Stability (ESC) | Hoye (2011) | Meta-analysis update; loss-of-control, single-vehicle, multi-vehicle | Crash occurrence | Loss-of-control about -40%; single-vehicle -25%; all fatal -40%; non-fatal overall near zero; multi-vehicle -6% to -23%, mostly non-significant |
| Stability (ESC) | Sivinski (2011) | United States, FARS and state data risk ratios | Fatal crash occurrence | Fatal involvement -23% (cars), -20% (light trucks and vans); fatal rollover -56% to -74% |
| Forward (FCW/AEB) | Kusano and Gabler (2012) | United States, simulation of 1,396 reconstructed tow-away rear-end crashes | Severity mitigation given a crash, plus occurrence | Median impact speed change -14% to -34%; moderately to fatally injured belted drivers -29% to -50%; crashes prevented 3.2% to 7.7% |
| Forward (FCW/AEB) | Cicchino (2017) | United States, 22 states, Poisson regression on police-reported crashes | Crash occurrence | Front-to-rear striking -27% (FCW), -43% (AEB), -50% (both); with injuries -20%, -45%, -56%; with third-party injuries -18% (marginally significant), -44%, -59% |
| Forward (AEB) | Cicchino and Zuby (2019) | United States, residual rear-end strikes by AEB-equipped vehicles | Residual-crash composition | Over-represented in snow or ice (odds ratio 1.83) and on roads with speed limits of 70 mph or more (odds ratio 1.49) |
| Forward (AEB) | Hu and Cicchino (2025) | United States, 19 states, 2017 to 2023; VIN-decoded striking-vehicle AEB; Poisson regression of rear-end crashes at signalized intersections | Crash occurrence | Rear-end crashes -36% (CI -39, -32); rear-end injury crashes -41% (CI -47, -33) |
| Lateral (LDW) | Kusano et al. (2014) | United States, fleet simulation of departure scenarios | Occurrence and serious injury | Drift-out-of-lane crashes -28.9%; seriously injured drivers -24.3% |
| Lateral and stability (LDW, RSC) | Hickman et al. (2015) | United States, large trucks from 14 trucking companies; Poisson regression on company crash records | Crash occurrence | LDW-related crashes -48% (CI -65, -23); rollover crashes -36% (CI -47, -22) |
| Lateral (LDW) | Sternlund et al. (2017) | Sweden, induced exposure; head-on and single-vehicle injury crashes | Injury-crash occurrence | -30% overall (95% CI lower limit 6%); -53% on 70 to 120 km/h roads with visible road markings (lower |

| Function group | Study | Data and design | Margin evaluated | Principal estimates |
|---|---|---|---|---|
| | | | | limit 11%) |
| Lateral (LDW) | Cicchino (2018a) | United States, Poisson regression on police-reported crashes | Crash occurrence | Relevant crash types -11% adjusted (-18% unadjusted); injury crashes of those types -21% adjusted, marginally significant (-24% unadjusted) |
| Lateral (BSW) | Cicchino (2018b) | United States, 26 states, 3.4 million insured vehicle years; same-direction lane-change crashes | Crash occurrence | Lane-change crash involvement -14% (CI -24, -2); injury lane-change crashes -23% (CI -44, +6), not significant |
| Lateral (LKA with LDW) | Scharber (2024) | United States, FARS 2016 to 2022; quasi-induced exposure, mixed-effects logistic regression by crash configuration | Fatal crash occurrence | Fatal single-vehicle road departure -24% (CI -42, -2); fatal same-direction sideswipe -1% (CI -34, +47), not significant; fatal opposite-direction -33% (CI -49, -11) |
| Forward and lateral (FCW, BSW) | Ding et al. (2025) | United States (Florida), 2017 to 2023; two-vehicle crashes; quasi-induced exposure comparing equipped and unequipped vehicles | Crash occurrence | Rear-end crashes -38% (FCW) and lane-change crashes -27% (BSW) in clear weather; non-incapacitating injury crashes -44% and incapacitating injury crashes -38% (BSW) |
| Forward and lateral (AEB, BSW) | IIHS and HLDI (2025) | United States, passenger vehicles; crash and insurance claim rates, equipped versus unequipped vehicles | Crash and claim occurrence | Front-to-rear crashes -50% and injury claims for people in other vehicles -24% (AEB); lane-change crashes -14% and injury claims for people in other vehicles -8% (BSW) |
| Forward and lateral (AEB, PAEB, LDW/LKA/LCA) | PARTS (2025a) | United States, 16 states, 2016 to 2023; 98 million model year 2015 to 2023 vehicles linked to crashes by VIN; quasi-induced exposure | Crash occurrence | Front-to-rear striking -49% (CI -50, -48), rising from -46% (2015 to 2017 models) to -52% (2021 to 2023 models); PAEB frontal non-motorist -9%; road departure -5% (LDW with LKA); LDW alone not significant |
| Forward and lateral (AEB, PAEB, LDW/LKA/LCA) | PARTS (2025b) | United States, 16 states, 2016 to 2023; 98 million model year 2015 to 2023 vehicles linked to injury crashes by VIN; quasi-induced exposure | Injury-crash occurrence | Front-to-rear striking -49% overall, -55% for moderate or worse injury, -42% for fatal or serious injury; PAEB non-motorist injury crashes -8.5% |

Note: ESC = electronic stability control; FCW = forward collision warning; AEB = automatic emergency braking; PAEB = pedestrian automatic emergency braking; LDW = lane departure warning; LKA = lane keeping assistance; LCA = lane centering assistance; BSW = blind spot warning; CI = confidence interval; SUV = sport utility vehicle; VIN = vehicle identification number; FARS = Fatality Analysis Reporting System; PARTS = Partnership for Analytics Research in Traffic Safety; RSC = roll stability control; IIHS = Insurance Institute for Highway Safety; HLDI = Highway Loss Data Institute.

The lateral functions, which act most directly on sideswipes, show more modest effects. Lane departure warning reduced relevant police-reported crash rates in the United States, with larger estimates from Sweden and from simulation (Cicchino 2018a; Sternlund et al. 2017; Kusano et al. 2014). It also reduced the crashes it targets in heavy-truck fleets (Hickman et al. 2015). Blind spot warning reduced lane-change crashes (Cicchino 2018b), and vehicles equipped with it or with forward collision warning had fewer two-vehicle crashes under most conditions during the maneuver each function supports (Ding et al. 2025). Few evaluations, however, isolate the same-direction sideswipe, and their estimates range from no clear effect in fatal crashes (Scharber 2024) to reductions with lane keeping assistance in police-reported crashes (Leslie et al. 2026). These evaluations also have two limits. Drivers do not always keep warning and assistance functions turned on (Cicchino 2018a; Kim and Oviedo-Trespalacios 2025), and equipped vehicles tend to be newer, so comparisons with unequipped vehicles may partly reflect vehicle age (Zhang et al. 2025).

The crash-prevention benefits of ADAS functions also reach people in other vehicles, through fewer rear-end crashes that injure them (Cicchino 2017) and lower injury claim rates (IIHS and HLDI 2025). This record says little about the severity margin. A severity sample includes only crashes that occurred (Abdel-Aty and Abdelwahab 2004c), so the crashes these functions prevent are not in it, and such events are better captured by surrogate safety measures of near-crashes, which do not depend on a crash having occurred (Anowar et al. 2025). A study that did model severity covered only vehicles with SAE Level 2 and Level 4 automation and did not compare them with unequipped vehicles (Ding et al. 2024). How function availability relates to the injury severity of both the striking and the struck driver therefore has rarely been measured in real-world crash data.

### *2.3. Joint models of paired injury outcomes*

Treating the two outcomes of one crash as unrelated equations discards information, since modeling them together gives more precise estimates and can change the results (Ouyang et al. 2002). Joint models have therefore been applied to a driver and a passenger of the same vehicle (Yamamoto and Shankar 2004) and to the two drivers of a crash (Abay et al. 2013). Other joint models have separated the at-fault and not-at-fault drivers in signalized-intersection crashes (Chiou et al. 2013), angle collisions (Russo et al. 2014), head-on collisions (Wali et al. 2018), and intersection-related rear-end crashes (Russo et al. 2023). The positive correlations between the two drivers' equations (Russo et al. 2014, 2023) indicate that their injuries share unobserved crash factors, which supports joint estimation.

These models have also been extended to address unobserved heterogeneity and temporal instability. Because fixed parameters can lead to biased estimates and erroneous inferences when the effects of observed factors vary across crashes (Mannering et al. 2016), random parameters have been widely adopted (Fountas et al. 2018; Behnood and Al-Bdairi 2020). Temporal instability has also received attention, as pooling four or more years of rear-end crash data generally yields structurally unstable models (Dabbour et al. 2020). The most fully specified of these joint models is a correlated joint random parameters bivariate probit of the following and leading drivers' injuries in freeway rear-end crashes, which also tests for temporal instability (Wang et al. 2024b).

The reviewed evidence therefore leaves a single gap. Severity evidence by driver role remains thin and concentrated in rear-end crashes, and the ADAS functions have been evaluated mostly at the crash-occurrence margin. What is needed is a severity analysis of real-world two-vehicle rear-end and same-direction sideswipe crashes that models the striking and struck drivers jointly and includes the individual ADAS functions of both vehicles.

## 3. Data description

### *3.1. Crash data and sample formation*

The crash data are drawn from the Crash Records Information System (CRIS) maintained by the Texas Department of Transportation, which assembles the police crash reports filed across the state (TxDOT 2025). The analysis covers the two-vehicle rear-end and sideswipe crashes recorded statewide from 2023 to 2025. Both crash types are identified from the reported manner of collision, and both are restricted to same-direction collisions. Rear-end crashes are those in which a vehicle strikes a slowing, stopped, or moving lead vehicle from behind. Sideswipe crashes are those in which two vehicles traveling in the same direction make lateral contact, including sideswipes initiated by turning movements. The two samples therefore share a common traffic setting and differ in the collision mechanism, and the two crash types are modeled separately on this basis.

Each crash pairs an at-fault striking vehicle with a struck vehicle, and the two roles were assigned from the contributing actions, violations, and driver fault recorded for each vehicle on the crash report. Only crashes in which this evidence clearly identified one striking and one struck vehicle were retained, so that every pair has an unambiguous role assignment. Because the assignment rests on the reporting officer's coding of those actions, some crashes could have a role misclassification that the data cannot reveal. Misclassification of this kind, if present, would mix the two roles and blur the contrast between the drivers, attenuating the estimated differences between them rather than inflating them. The reported associations are conservative in this respect. Joining the two vehicle records of each crash into one paired observation yields 2,461 rear-end and 1,437 sideswipe crashes.

The injury outcome follows the KABCO convention of the police report, with recorded levels fatal injury (K), suspected serious injury (A), suspected minor injury (B), possible injury (C), and no apparent injury (O) (TxDOT 2023; NHTSA 2017). Fatal and suspected serious injuries together account for at most 0.7% of either vehicle's outcomes in these crash types. The analysis therefore adopts a binary outcome contrasting any reported injury (levels K, A, B, or C) with no reported injury (level O), because models with several severity levels perform poorly when the severity levels are imbalanced (Anowar 2026).

The estimation samples restrict both vehicles to the two light-vehicle body classes that dominate the pairs, passenger cars and a combined class of light trucks, vans, and sport utility vehicles. They also require complete information on the driver, vehicle, and roadway attributes used in the model specification process. A crash pair was excluded if either vehicle was of another body type, such as a heavy truck, bus, motorcycle, emergency vehicle, or unclassified body style. A pair was also excluded if either driver's age or sex, or the posted speed limit, was missing. These restrictions leave 2,273 rear-end and 1,216 sideswipe crashes. Table 2 reports the severity distributions of the two vehicles in both estimation samples. Rear-end crashes are more injurious than sideswipe crashes overall, and the struck vehicle has the higher injured share in both. These patterns are consistent with the rear impact transferring collision energy into the lead vehicle, while same-direction lateral contact is more often a lighter sideways impact.

**Table 2 Injury severity of the striking and struck vehicles in the estimation samples**

| Severity level | Rear-end crashes | | Sideswipe crashes | |
|---|---|---|---|---|
| | Striking | Struck | Striking | Struck |
| Fatal injury (K) | 4 (0.2%) | 2 (0.1%) | 1 (0.1%) | 0 (0.0%) |
| Suspected serious injury (A) | 12 (0.5%) | 5 (0.2%) | 1 (0.1%) | 7 (0.6%) |
| Suspected minor injury (B) | 134 (5.9%) | 169 (7.4%) | 40 (3.3%) | 51 (4.2%) |
| Possible injury (C) | 141 (6.2%) | 308 (13.6%) | 57 (4.7%) | 94 (7.7%) |
| No apparent injury (O) | 1,982 (87.2%) | 1,789 (78.7%) | 1,117 (91.9%) | 1,064 (87.5%) |
| Injury (K, A, B, C) | 291 (12.8%) | 484 (21.3%) | 99 (8.1%) | 152 (12.5%) |
| Total crashes | 2,273 | | 1,216 | |

*3.2. ADAS function classification*

For every vehicle in the sample, the availability of thirteen advanced driver assistance system (ADAS) functions was established from the vehicle identification number (VIN) recorded on the crash report. The stability functions include electronic stability control (ESC), traction control (TC), and anti-lock braking (ABS), whereas the forward-looking functions include forward collision warning (FCW), crash imminent braking (CIB), and adaptive cruise control (ACC). The lateral and remaining functions are lane departure warning (LDW), lane keeping assistance (LKA), lane centering assistance (LCA), blind spot warning (BSW), parking assist (PA), pedestrian automatic emergency braking (PAEB), and rear automatic emergency braking (RAEB). Each VIN was decoded through the National Highway Traffic Safety Administration's Product Information Catalog and Vehicle Listing (vPIC), which reports equipment availability at the make, model, and model year level (NHTSA 2026).

Each function on each vehicle is coded as confirmed standard, optional, confirmed not available, or unknown. Entries that vPIC leaves unknown are filled from independent public sources in a fixed order that never overwrites confirmed information. New Car Assessment Program safety ratings supply ESC, FCW, and LDW. The Federal Motor Vehicle Safety Standard No. 126 requirement made ESC, together with the ABS and TC capability it builds on, standard on all new light vehicles by model year 2012 (NHTSA 2007). Where the crash form recorded that a driving automation system was engaged, the reported automation level implies ACC and LCA under the SAE J3016 taxonomy (SAE International 2021). Only 2.5% of the rear-end vehicles and 3.2% of the sideswipe vehicles have no resolved function at all, almost all of them older vehicles, reflecting the sparseness of vPIC equipment records before model year 2017. A function is counted as present only when it is standard equipment on the vehicle.

*3.3. Crash involvement ratio*

Beyond the ADAS function indicators, the models draw on the driver, vehicle, and roadway attributes recorded on the crash report, covering the drivers' demographics and behavior, the vehicles' body class and vehicle age, and the roadway and traffic context of the crash. Vehicle age is the crash year minus the vehicle's model year, grouped into four categories (under 5 years, 5 to 9 years, 10 to 15 years, and 16 years or more).

Each crash location is further characterized by the composition of its street-level visual scene. Two Google Street View images retrieved at the crash coordinates were processed with a transformer-based semantic segmentation model, following the approach applied to intersection crash analysis by Wang et al. (2025). The share of image pixels falling in each of nineteen scene classes describes the visual environment of the crash site. These shares are crash-level attributes common to both vehicles of a pair. Because archived imagery need not coincide with the crash date or its momentary lighting and weather, the shares are interpreted as indicators of the built environment of the site rather than of conditions at the moment of the crash (Wang et al. 2025).

All continuous variables, including the scene shares, enter the models in standardized (z-scored) form, and each categorical variable enters as a set of indicator variables, with its reference category stated in the model results tables. Table A1 in Appendix A reports the descriptive statistics of the model variables for both crash types.

Before modeling, the striking and struck vehicles are compared using the crash involvement ratio (CIR), a quasi-induced exposure measure (Wang et al. 2024a; Zhang et al. 2024). The method assumes that drivers who are not at fault reflect typical driving exposure, so the struck vehicles are treated as the control group. For each category of a driver or vehicle attribute, the CIR is calculated as the striking vehicles' share in that category divided by the struck vehicles' share, as shown in Eq. (1).

$$\mathrm{CIR}_i = \frac{\frac{n_{1i}}{\sum_i n_{1i}}}{\frac{n_{2i}}{\sum_i n_{2i}}} \tag{1}$$

Here, $\mathrm{CIR}_i$ is the crash involvement ratio of category $i$, $n_{1i}$ and $n_{2i}$ are the numbers of striking and struck vehicles in category $i$, and $\sum_i n_{1i}$ and $\sum_i n_{2i}$ are the total numbers of striking and struck vehicles. A CIR above 1 therefore means that the category is more common among striking vehicles, while a value below 1 means that it is less common. Crash-level attributes, such as the posted speed limit and the scene shares, are not included because they take the same value for both vehicles.

Table 3 presents the ratios for both crash types, and Figure 1 illustrates the ratios for selected categories. The table covers the ADAS functions used in the models, with ESC, ABS, and TC for both crash types and LDW and LCA for sideswipe crashes only. In both crash types, drivers under 25 years and unlicensed drivers are over-represented in the striking role, and so are vehicles 10 years or older.

Other patterns, however, depend on the crash type. Male drivers are over-represented in the striking role mainly in rear-end crashes (1.26, against 1.02 in sideswipe crashes), whereas drivers 65 years or older are over-represented only in sideswipe crashes. In addition, vehicles without the ADAS functions are over-represented in the striking role, particularly in rear-end crashes. This is largely because these vehicles tend to be older, as the vehicle age ratios suggest. The ratios are descriptive, with no control for other attributes and no confidence intervals, so they should not be interpreted as causal effects. Still, they show that the striking and struck vehicles differ in composition. This difference supports estimating each driver's injury in its own equation with its own set of variables.

**Table 3 Crash involvement ratios for driver, vehicle, and ADAS characteristics**

| Categories | Subgroup | Rear-end (N = 2,273) | | | Sideswipe (N = 1,216) | | |
|---|---|---|---|---|---|---|---|
| | | Striking n (%) | Struck n (%) | CIR | Striking n (%) | Struck n (%) | CIR |
| Driver characteristics | Male | 1,367 (60.1%) | 1,082 (47.6%) | **1.26** | 664 (54.6%) | 648 (53.3%) | **1.02** |
| | Female | 906 (39.9%) | 1,191 (52.4%) | 0.76 | 552 (45.4%) | 568 (46.7%) | 0.97 |
| | Under 25 years | 692 (30.4%) | 344 (15.1%) | **2.01** | 303 (24.9%) | 197 (16.2%) | **1.54** |
| | 25 to 64 years | 1,424 (62.6%) | 1,736 (76.4%) | 0.82 | 773 (63.6%) | 913 (75.1%) | 0.85 |
| | 65 years or over | 157 (6.9%) | 193 (8.5%) | 0.81 | 140 (11.5%) | 106 (8.7%) | **1.32** |
| | Licensed | 1,998 (87.9%) | 2,128 (93.6%) | 0.94 | 1,076 (88.5%) | 1,137 (93.5%) | 0.95 |
| | Not licensed | 275 (12.1%) | 145 (6.4%) | **1.90** | 140 (11.5%) | 79 (6.5%) | **1.77** |
| | One occupant | 1,778 (78.2%) | 1,535 (67.5%) | **1.16** | 890 (73.2%) | 848 (69.7%) | **1.05** |
| | Two occupants | 324 (14.3%) | 485 (21.3%) | 0.67 | 227 (18.7%) | 246 (20.2%) | 0.92 |
| | Three or more occupants | 171 (7.5%) | 253 (11.1%) | 0.68 | 99 (8.1%) | 122 (10.0%) | 0.81 |
| Vehicle characteristics | Larger body (light truck, van, or SUV) | 1,246 (54.8%) | 1,318 (58.0%) | 0.95 | 675 (55.5%) | 678 (55.8%) | 1.00 |
| | Passenger car | 1,027 (45.2%) | 955 (42.0%) | **1.08** | 541 (44.5%) | 538 (44.2%) | **1.01** |
| | Vehicle age under 5 years | 628 (27.6%) | 1,318 (58.0%) | 0.48 | 503 (41.4%) | 628 (51.6%) | 0.80 |
| | Vehicle age 5 to 9 years | 671 (29.5%) | 545 (24.0%) | **1.23** | 332 (27.3%) | 347 (28.5%) | 0.96 |
| | Vehicle age 10 to 15 years | 520 (22.9%) | 256 (11.3%) | **2.03** | 217 (17.8%) | 140 (11.5%) | **1.55** |
| | Vehicle age 16 years or more | 454 (20.0%) | 154 (6.8%) | **2.95** | 164 (13.5%) | 101 (8.3%) | **1.62** |
| ADAS functions | ESC available | 1,839 (80.9%) | 2,110 (92.8%) | 0.87 | 1,120 (92.1%) | 1,146 (94.2%) | 0.98 |
| | ESC not available | 434 (19.1%) | 163 (7.2%) | **2.66** | 96 (7.9%) | 70 (5.8%) | **1.37** |
| | ABS available | 1,636 (72.0%) | 2,035 (89.5%) | 0.80 | 986 (81.1%) | 1,066 (87.7%) | 0.92 |
| | ABS not available | 637 (28.0%) | 238 (10.5%) | **2.68** | 230 (18.9%) | 150 (12.3%) | **1.53** |
| | TC available | 1,616 (71.1%) | 2,029 (89.3%) | 0.80 | 976 (80.3%) | 1,062 (87.3%) | 0.92 |
| | TC not available | 657 (28.9%) | 244 (10.7%) | **2.69** | 240 (19.7%) | 154 (12.7%) | **1.56** |
| | LDW available | - | - | - | 749 (61.6%) | 858 (70.6%) | 0.87 |
| | LDW not available | - | - | - | 467 (38.4%) | 358 (29.4%) | **1.30** |
| | LCA available | - | - | - | 239 (19.7%) | 289 (23.8%) | 0.83 |
| | LCA not available | - | - | - | 977 (80.3%) | 927 (76.2%) | **1.05** |

Note: CIR values above 1 are shown in bold.

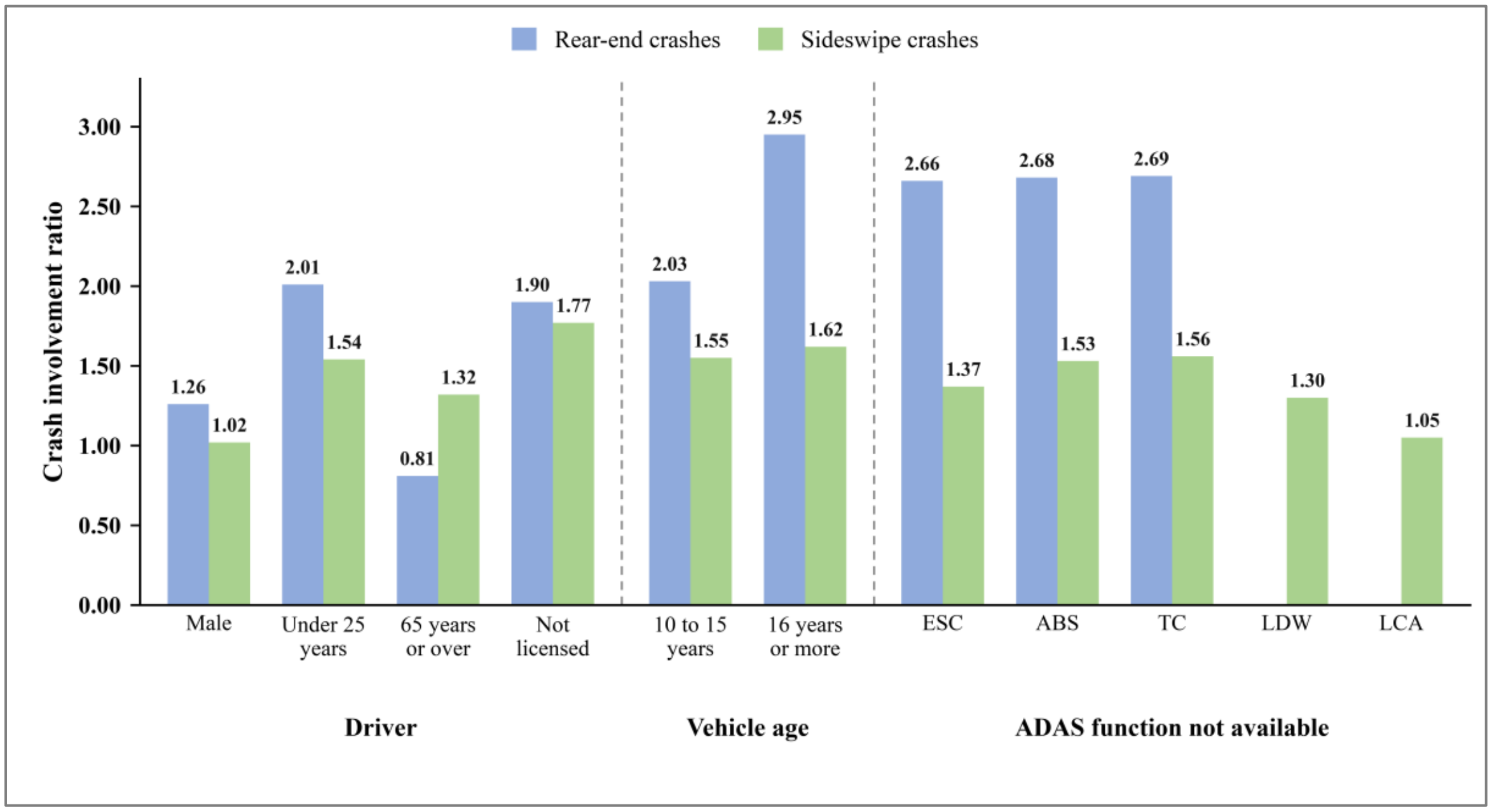

**Figure 1 Crash involvement ratios for selected driver, vehicle, and ADAS function categories**

## 4. Methodology

In a two-vehicle crash, the injury outcomes of the two vehicles arise from the same physical event. They are therefore exposed to common unmeasured factors, such as the impact configuration, the collision energy, and conditions at the crash site that the observed variables do not capture. Modeling the two outcomes with independent univariate equations ignores this dependence. Separate equations produce less precise estimates, provide no measure of how strongly the two outcomes are related, and misstate the probability of each combination of outcomes (Washington et al. 2020). Accordingly, this study represents the injury outcomes of the striking and struck vehicles as two latent propensity equations tied together by a correlation between their error terms, an approach applied to the paired injury outcomes of two-vehicle rear-end crashes (Wang et al. 2024b). The striking and struck roles are assigned as described in Section 3.1, and the framework is applied separately to rear-end and sideswipe crashes. A random parameters extension of the base specification allows selected coefficients to differ across crashes, capturing unobserved heterogeneity in the effects of observed factors (Mannering et al. 2016).

### *4.1. Joint bivariate probit model*

The joint model consists of two binary probit equations, one for each vehicle, whose latent propensities determine the observed outcomes (Washington et al. 2020),

$$\begin{aligned} S_{i1} &= \beta_1' X_{i1} + \varepsilon_{i1}, \; s_{i1} = 1 \text{ if } S_{i1} > 0, 0 \text{ otherwise} \\ S_{i2} &= \beta_2' X_{i2} + \varepsilon_{i2}, \; s_{i2} = 1 \text{ if } S_{i2} > 0, 0 \text{ otherwise} \end{aligned} \tag{2}$$

Here, $i$ $(i = 1, \ldots, N)$ indexes two-vehicle crashes; the subscripts 1 and 2 denote the striking vehicle and the struck vehicle, respectively; $S_{i1}$ and $S_{i2}$ are latent injury propensities; and the binary outcomes $s_{i1}$ and $s_{i2}$ equal 1 if injury (level K, A, B, or C of the KABCO scale) was reported for the corresponding vehicle and 0 if no injury was reported (level O), as detailed in Section 3.1. Each equation has its own explanatory variables, collected in $X_{i1}$ and $X_{i2}$, with $\beta_1$ and $\beta_2$ the parameter vectors to be estimated. The two covariate vectors can differ, and each may contain attributes of both vehicles and both drivers. Within the same model, a vehicle's ADAS functions can therefore be related to the injury outcome in that vehicle and in the other vehicle. The covariate set of each equation is selected independently for each crash type, and the statistically significant estimates are reported with the empirical results. The error terms are assumed to follow a standardized bivariate normal distribution,

$$(\varepsilon_{i1}, \varepsilon_{i2})' \sim \mathrm{N}\left((0,0)', \begin{matrix} 1 & \rho \\ \rho & 1 \end{matrix}\right) \quad (3)$$

Here, both error terms are standardized to zero mean and unit variance, and $\rho$ measures how strongly the unexplained components of the two equations move together. A positive $\rho$ indicates that unmeasured factors increasing the injury propensity of one vehicle also tend to increase the injury propensity of the other vehicle in the same crash. The joint density of the standardized bivariate normal takes the form (Washington et al. 2020),

$$\varphi_2(z_1, z_2;\ \rho) \ = \ \frac{\exp\left(-\frac{z_1{}^2 + z_2{}^2 - 2\rho\, z_1\, z_2}{2(1-\rho^2)}\right)}{2\pi\sqrt{1-\rho^2}} \quad (4)$$

Integrating this density over the region below the two evaluation points gives the probability that an injury occurs in both vehicles,

$$P(s_{i1} \ = \ 1, s_{i2} \ = \ 1) \ = \ \Phi_2(\beta_1'\, X_{i1}, \beta_2'\, X_{i2};\ \rho) \quad (5)$$

Here, $\Phi_2$ denotes the cumulative distribution function of the standardized bivariate normal distribution. The probabilities of the remaining three outcome combinations follow from the symmetry of the bivariate normal distribution and can be written compactly by defining the sign variables $q_{i1} = 2\, s_{i1} - 1$ and $q_{i2} = 2\, s_{i2} - 1$, giving

$$P(s_{i1}, s_{i2}) \ = \ \Phi_2(q_{i1}(\beta_1'\, X_{i1}), q_{i2}(\beta_2'\, X_{i2});\ q_{i1}\, q_{i2}\, \rho) \quad (6)$$

This single expression covers all four combinations of injury and no injury across the pair. Summing the log of Eq. (6) over the sample yields the log-likelihood that maximum likelihood estimation maximizes (Washington et al. 2020),

$$LL \ = \ \textstyle\sum_i \ln \Phi_2(q_{i1}(\beta_1'\, X_{i1}), q_{i2}(\beta_2'\, X_{i2});\ q_{i1}\, q_{i2}\, \rho) \quad (7)$$

This log-likelihood is equivalent to the four-term representation in which each outcome pair contributes its own bivariate normal probability.

*4.2. Random parameters extension*

Estimated parameters that are restricted to be constant across crashes can mask systematic variation in the effects of observed factors that arises from unobserved characteristics of drivers, vehicles, and crash circumstances (Mannering et al. 2016). To account for this unobserved heterogeneity, selected elements of the parameter vector of each equation are allowed to vary across crashes,

$$\beta_{ij} \ = \ \beta_j \ + \ \sigma_j\, \nu_{ij}\ ,\ j \ = \ 1, 2 \quad (8)$$

Here, for equation $j$, $\beta_j$ is the mean vector of the parameter distributions, $\sigma_j$ is a diagonal matrix of estimable standard deviation parameters, and $\nu_{ij}$ is a vector of independent standard normal disturbances. A parameter whose standard deviation is not statistically distinguishable from zero is restricted to be fixed across observations. All random parameters are specified as normally distributed, the density that injury severity applications have repeatedly found to fit best (Milton et al. 2008; Behnood and Al-Bdairi 2020). Under Eq. (8), the outcome-pair probability in Eq. (6) becomes

$$P_i(s_{i1}, s_{i2}) \ = \ \int \Phi_2(q_{i1}(\beta_{i1}'\, X_{i1}), q_{i2}(\beta_{i2}'\, X_{i2});\ q_{i1}\, q_{i2}\, \rho)\, f(\nu_{i1})\, f(\nu_{i2})\, \mathrm{d}\, \nu_{i1}\, \mathrm{d}\, \nu_{i2} \quad (9)$$

Here, $f(\cdot)$ denotes the product of independent standard normal densities of the random parameter disturbances, and the integral is taken over their distribution.

*4.3. Model estimation and marginal effects*

The fixed parameters specification is estimated by maximum likelihood, maximizing the log-likelihood in Eq. (7). Because the integral in Eq. (9) has no closed form, the random parameters models are estimated by simulated maximum likelihood (McFadden and Train 2000). The integral is approximated with 1,000 Halton draws, a number that prior severity studies found sufficient at acceptable computational cost (Behnood and Al-Bdairi 2020; Wang et al. 2024b).

The fixed parameters model is nested within the random parameters model, arising when all standard deviation parameters in Eq. (8) are restricted to zero. Competing specifications are therefore compared with nested likelihood ratio tests, in which twice the difference between the log-likelihoods at convergence is chi-square distributed with degrees of freedom equal to the number of additional parameters. Model fit is further assessed with the log-likelihood at convergence together with the Akaike information criterion (AIC) and the Bayesian information criterion (BIC), which penalize parameterization. Triangular and uniform distributions are also tested for the two random parameters of the rear-end model (Table A2). With the same number of parameters, their log-likelihoods at convergence (-1871.65 and -1871.61) are practically the same as that of the normal specification (-1871.67). Because the likelihood ratio test cannot compare these forms with the normal one, they are compared with it using the Vuong (1989) test. The test does not distinguish either of them from the normal specification (z = -0.19 and -0.21, against a critical value of ±1.96 at the 95% confidence level), and the ADAS function estimates remain essentially unchanged, so the normal distribution is retained.

Probit coefficients do not have a direct probability interpretation, so the estimates are translated onto the probability scale through marginal effects, computed with all other variables held at their observed values. Let $x_k$ denote an indicator variable contained in $X_{i1}$ or $X_{i2}$. The marginal effect of $x_k$ is computed as the discrete change in the outcome probability as $x_k$ moves from zero to one (Christofides et al. 1997; Greene 2012),

$$E_{x_k} = P(s_{i1}, s_{i2} \mid x_k = 1) - P(s_{i1}, s_{i2} \mid x_k = 0) \quad (10)$$

This change is evaluated for each crash and averaged across all $N$ observations, with the corresponding expressions applying to each vehicle's individual injury probability. For a z-scored continuous variable, the marginal effect is computed analogously, as the change in the outcome probability when the variable increases by one standard deviation (Greene 2012).

## 5. Results and discussion

### *5.1. Model specification and overall measure of fit*

Table 4 presents the estimation results and marginal effects for the rear-end model and Table 5 for the sideswipe model, each reporting the striking-driver and struck-driver results side by side with the cross-equation correlation and the fit statistics. For each crash type, a fixed parameters specification and a random parameters specification are estimated. For rear-end crashes, the likelihood ratio test comparing the two forms gives 5.86 with two degrees of freedom, which is significant at the 90% confidence level. For sideswipe crashes, random parameters were examined and none was retained. Therefore, the random parameters specification is adopted for rear-end crashes and the fixed parameters specification for sideswipe crashes. A single model is estimated for each crash type over the full study period, and a likelihood ratio test finds no evidence against pooling the study years for either rear-end or sideswipe crashes.

**Table 4 Model estimation results and marginal effects for two-vehicle rear-end crashes based on the random parameters joint bivariate probit model**

| Variable | Model estimates (t-stat) | | Marginal effects (for injury) | |
|---|---|---|---|---|
| | V1 driver | V2 driver | V1 driver | V2 driver |
| **Driver characteristics** | | | | |
| V1 male driver (1 if male, 0 otherwise) | -0.220 (-2.815) | 0.178 (2.550) | -0.0390 | 0.0446 |
| V2 male driver (1 if male, 0 otherwise) | - | -0.223 (-3.375) | - | -0.0566 |
| V1 driver < 25 (1 if age is < 25 years, 0 otherwise) | -0.161 (-1.882) | - | -0.0272 | - |
| V1 driver unrestrained (1 if not restrained, 0 otherwise) | 1.096 (3.105) | - | 0.2821 | - |
| V2 driver unrestrained (1 if not restrained, 0 otherwise) | - | 1.428 (1.991) | - | 0.4643 |
| V1 driver impaired (1 if impaired, 0 otherwise) | 0.527 (1.971) | - | 0.1141 | - |
| V1 two occupants (1 if two persons in the vehicle, | 0.398 (3.945) | - | 0.0785 | - |

| Variable | Model estimates (t-stat) | | Marginal effects (for injury) | |
|---|---|---|---|---|
| | V1 driver | V2 driver | V1 driver | V2 driver |
| 0 otherwise) | | | | |
| V1 three or more occupants (1 if three or more persons in the vehicle, 0 otherwise) | 0.438 (3.298) | - | 0.0898 | - |
| V2 two occupants (1 if two persons in the vehicle, 0 otherwise) | - | 0.260 (3.345) | - | 0.0696 |
| V2 three or more occupants (1 if three or more persons in the vehicle, 0 otherwise) | - | 0.390 (3.937) | - | 0.1093 |
| **Vehicle characteristics** | | | | |
| V1 larger body (1 if a light truck, van, or SUV, 0 if a passenger car) | -0.322 (-4.062) | 0.209 (3.069) | -0.0567 | 0.0529 |
| V2 larger body (1 if a light truck, van, or SUV, 0 if a passenger car) | 0.161 (2.021) | -0.174 (-2.598) | 0.0277 | -0.0447 |
| **Roadway and contextual characteristics** | | | | |
| Posted speed limit (mph, z-scored) | 0.236 (5.163) | 0.113 (2.583) | 0.0456 | 0.0299 |
| Interstate roadway (1 if an interstate, 0 otherwise) | -0.283 (-2.918) | -0.149 (-1.701) | -0.0462 | -0.0369 |
| Local roadway (1 if a local road, 0 otherwise) | - | 0.237 (2.712) | - | 0.0624 |
| Service road (1 if a service road, 0 otherwise) | - | 0.446 (3.637) | - | 0.1274 |
| Traffic control signalized (1 if signalized, 0 otherwise) | -0.201 (-2.046) | - | -0.0332 | - |
| Traffic control none or uncontrolled (1 if no traffic control, 0 otherwise) | -0.386 (-2.075) | - | -0.0564 | - |
| Crash time night (1 if night period, 0 otherwise) | -0.243 (-1.707) | - | -0.0390 | - |
| Weekend (1 if Saturday or Sunday, 0 otherwise) | 0.172 (1.861) | - | 0.0315 | - |
| Dark, lighted (1 if dark but lighted, 0 otherwise) | 0.282 (1.838) | - | 0.0542 | - |
| Building density (share of scene pixels, z-scored) | 0.068 (1.825) | 0.058 (1.884) | 0.0121 | 0.0151 |
| **Automation (ADAS functions)** | | | | |
| V1 ESC without ABS (1 if ESC is standard and ABS is not, 0 otherwise; base: no ESC) | - | -0.317 (-2.405) | - | -0.0731 |
| V1 ESC and ABS (1 if ESC and ABS are standard, 0 otherwise; base: no ESC) | - | -0.139 (-1.685) | - | -0.0360 |
| V2 TC and ABS (1 if TC and ABS are standard, 0 otherwise) | - | -0.348 (-3.377) | - | -0.0967 |
| **Constant** | | | | |
| Constant | -0.853 (-7.770) | -0.639 (-4.140) | - | - |
| **Random parameters** | | | | |
| Crash time peak (1 if the morning or evening peak period, 0 otherwise) | -0.567 (-2.458) | - | -0.0420 | - |
| ***Standard deviation of crash time peak*** | ***0.820 (2.992)*** | - | - | - |
| Intersection context (1 if intersection or driveway related, 0 otherwise) | - | -0.039 (-0.271) | - | -0.0101 |
| ***Standard deviation of intersection context*** | - | ***0.690 (2.323)*** | - | - |
| **Cross-equation correlation** | | | | |
| Cross-equation correlation rho | 0.559 (9.460) | | | |
| **Model statistics** | | | | |
| Number of observations (pairs) | 2,273 | | | |
| Log-likelihood at convergence, LL | -1871.7 | | | |
| Log-likelihood at zero, LL(0) | -1999.8 | | | |
| AIC | 3825.3 | | | |
| BIC | 4060.2 | | | |
| Number of parameters | 41 | | | |

Note: absolute t values above 1.645, 1.96, and 2.576 indicate statistical significance at the 90%, 95%, and 99% confidence levels. V1 = striking-vehicle, and V2 = struck-vehicle. Fixed parameters model: LL = -1874.6, LL(0) = -1999.8, AIC = 3827.2.

**Table 5 Model estimation results and marginal effects for two-vehicle sideswipe crashes based on the fixed parameters joint bivariate probit model**

| Variable | Model estimates (t-stat) | | Marginal effects (for injury) | |
|---|---|---|---|---|
| | V1 driver | V2 driver | V1 driver | V2 driver |
| **Driver characteristics** | | | | |
| V1 male driver (1 if male, 0 otherwise) | -0.417 (-3.821) | - | -0.0597 | - |
| V2 male driver (1 if male, 0 otherwise) | - | -0.280 (-2.843) | - | -0.0531 |
| V2 driver < 25 (1 if age is < 25 years, 0 otherwise) | - | -0.297 (-2.052) | - | -0.0500 |
| V2 driver unlicensed (1 if not licensed, 0 otherwise) | - | 0.331 (1.770) | - | 0.0718 |
| V1 two occupants (1 if two persons in the vehicle, 0 otherwise) | 0.277 (2.105) | - | 0.0437 | - |
| V1 three or more occupants (1 if three or more persons in the vehicle, 0 otherwise) | 0.326 (1.841) | - | 0.0547 | - |
| V2 two occupants (1 if two persons in the vehicle, 0 otherwise) | - | 0.350 (3.116) | - | 0.0731 |
| **Vehicle characteristics** | | | | |
| V1 larger body (1 if a light truck, van, or SUV, 0 if a passenger car) | -0.227 (-2.104) | - | -0.0325 | - |
| V2 larger body (1 if a light truck, van, or SUV, 0 if a passenger car) | - | -0.271 (-2.748) | - | -0.0519 |
| **Vehicle age bands** | | | | |
| V1 vehicle age under 5 years (1 if the striking vehicle is under 5 years old, 0 otherwise; base: 16 years or more) | 0.686 (2.244) | 0.699 (2.952) | 0.1083 | 0.1417 |
| V1 vehicle age 5 to 9 years (1 if the striking vehicle is 5 to 9 years old, 0 otherwise; base: 16 years or more) | 0.589 (1.918) | 0.553 (2.507) | 0.1008 | 0.1187 |
| V2 vehicle age under 5 years (1 if the struck vehicle is under 5 years old, 0 otherwise; base: 16 years or more) | - | 0.796 (2.414) | - | 0.1494 |
| **Roadway and contextual characteristics** | | | | |
| Posted speed limit (mph, z-scored) | 0.127 (2.178) | - | 0.0194 | - |
| Intersection context (1 if intersection or driveway related, 0 otherwise) | 0.211 (1.824) | - | 0.0308 | - |
| Roadway not straight (1 if the alignment is not straight, 0 otherwise) | -0.687 (-2.249) | - | -0.0643 | - |
| Three-leg intersection (1 if three approach legs, 0 otherwise) | - | 0.455 (2.983) | - | 0.1023 |
| Four-leg intersection (1 if four approach legs, 0 otherwise) | - | 0.282 (1.927) | - | 0.0587 |
| Adverse weather (1 if cloudy, rain, fog, snow or sleet, or severe wind, 0 otherwise) | - | 0.193 (1.742) | - | 0.0384 |
| Dark, not lighted (1 if dark and unlighted, 0 otherwise) | - | 0.586 (2.560) | - | 0.1423 |
| Crash time peak (1 if the morning or evening peak period, 0 otherwise) | - | -0.171 (-1.725) | - | -0.0317 |
| Person density (share of scene pixels, z-scored) | 0.076 (1.950) | - | 0.0113 | - |
| Terrain (share of scene pixels, z-scored) | - | 0.099 (1.734) | - | 0.0197 |
| Road surface density (share of scene pixels, z-scored) | - | 0.153 (2.422) | - | 0.0310 |
| **Automation (ADAS functions)** | | | | |
| V1 TC and ABS (1 if TC and ABS are standard, 0 | -0.536 (-2.154) | - | -0.0946 | - |

| Variable | Model estimates (t-stat) | | Marginal effects (for injury) | |
|---|---|---|---|---|
| | V1 driver | V2 driver | V1 driver | V2 driver |
| otherwise) | | | | |
| V1 ESC (1 if ESC is standard, 0 otherwise) | - | -0.451 (-2.028) | - | -0.1018 |
| V1 LDW without LCA (1 if LDW is standard and LCA is not, 0 otherwise; base: no LDW) | - | -0.421 (-2.948) | - | -0.0778 |
| V1 LDW and LCA (1 if LDW and LCA are standard, 0 otherwise; base: no LDW) | - | -0.501 (-2.679) | - | -0.0791 |
| V2 TC and ABS (1 if TC and ABS are standard, 0 otherwise) | - | -0.462 (-1.747) | - | -0.1032 |
| **Constant** | | | | |
| Constant | -1.308 (-6.785) | -1.010 (-3.592) | - | - |
| **Cross-equation correlation** | | | | |
| Cross-equation correlation rho | 0.375 (4.857) | | | |
| **Model statistics** | | | | |
| Number of observations (pairs) | 1,216 | | | |
| Log-likelihood at convergence, LL | -728.3 | | | |
| Log-likelihood at zero, LL(0) | -793.7 | | | |
| AIC | 1532.7 | | | |
| BIC | 1726.6 | | | |
| Number of parameters | 38 | | | |

Note: absolute t values above 1.645, 1.96, and 2.576 indicate statistical significance at the 90%, 95%, and 99% confidence levels. V1 = striking-vehicle, and V2 = struck-vehicle.

The cross-equation correlations are positive and statistically significant in both models. The rear-end correlation is 0.559 (t = 9.460) and the sideswipe correlation is 0.375 (t = 4.857). A positive correlation means that a crash injuring one driver more severely than its observed characteristics predict tends to injure the other driver more severely as well. Unobserved crash-level factors, plausibly the impact configuration, the collision energy, and unmeasured local conditions, therefore move both drivers' outcomes in the same direction, and estimating the two outcomes jointly is warranted (Wang et al. 2024b). Recent joint models of two-vehicle crashes report positive correlations of a similar size, 0.431 between at-fault and not-at-fault drivers in intersection-related rear-end crashes (Russo et al. 2023) and 0.612 between following and leading drivers in freeway rear-end crashes (Wang et al. 2024c). In truck-car crashes, the correlation between the two drivers ranges from 0.283 to 0.624 across the pooled and yearly models of Song et al. (2023). A weaker positive correlation, 0.147, has also been reported between striking and struck vehicles in 130 side-impact crashes at intersections (Yuan et al. 2020).

### *5.2. Model estimation results*

The estimation results are discussed by characteristic family, following the order of the tables and ending with the random parameters. In Table 4 and Table 5, each estimate is shown as the coefficient with its t-statistic in parentheses. Only statistically significant estimates are reported, except that a random parameter's mean is always reported together with its standard deviation.

The coefficients alone do not convey the size of the associations, so marginal effects are computed as described in Section 4.3 and reported beside the estimates in Table 4 and Table 5. Each value is the average change in the corresponding driver's own injury probability when a binary indicator moves from zero to one or a z-scored variable increases by one standard deviation. For example, in rear-end crashes, a one standard deviation higher posted speed limit is associated with a striking-driver injury probability higher by 0.0456. The remaining values can be interpreted in the same way. To put these values in context, struck drivers are injured more often than striking drivers, 21.3% against 12.8% in rear-end crashes and 12.5% against 8.1% in sideswipe crashes. The subsections below report the size of the main associations together with their direction and significance.

*5.2.1. Driver characteristics*

The associations of the driver characteristics are largely consistent with previous injury severity studies. Male drivers, for instance, have lower own-injury probabilities in both rear-end and sideswipe crashes, a pattern usually attributed to differences in fragility and in the reporting of minor injury (Bhowmik et al. 2026). Drivers under 25 likewise have lower own-injury probabilities, as striking drivers in rear-end crashes and as struck drivers in sideswipe crashes, which is often explained by physiological differences from older drivers (Russo et al. 2014). By contrast, alcohol or drug impairment is associated with a higher own-injury probability for striking drivers in rear-end crashes, and the same holds for unlicensed struck drivers in sideswipe crashes. Carrying passengers is also associated with a higher injury probability for the driver carrying them in both crash types, possibly because passengers can influence how the driver responds before and during a crash (Seraneeprakarn et al. 2017). Beyond these own-driver associations, a male striking driver in rear-end crashes is associated with a higher injury probability for the struck driver, which may reflect driving style ahead of the crash.

Regarding the size of these associations, unrestrained drivers show the largest marginal effects among the driver characteristics, with higher own-injury probabilities for both striking and struck drivers in rear-end crashes. However, these estimates rest on very few cases (17 striking and 5 struck drivers), so they are not interpreted. Among the remaining driver characteristics, the largest marginal effects in rear-end crashes are 0.1141 for an impaired striking driver and 0.1093 for a struck driver with three or more occupants in the vehicle. In sideswipe crashes, the driver characteristics show smaller marginal effects, with the largest being 0.0731 for a struck driver with two occupants in the vehicle.

*5.2.2. Vehicle characteristics*

The vehicle size associations are consistent with vehicle compatibility research. A larger striking vehicle (a light truck, van, or sport utility vehicle) is associated with a lower own-driver injury probability in both crash types, and in rear-end crashes also with a higher struck-driver injury probability. In rear-end crashes, the corresponding marginal effects indicate an own-driver injury probability lower by 0.0567 and a struck-driver injury probability higher by 0.0529. Similarly, a larger struck vehicle is associated with a lower own-driver injury probability in both crash types, and in rear-end crashes with a modestly higher striking-driver injury probability. This asymmetry reflects the familiar compatibility pattern, in which the driver of the larger vehicle faces a lower injury risk and the driver of the other vehicle a higher one (Abdel-Aty and Abdelwahab 2004a).

Vehicle age, however, shows a different pattern in sideswipe crashes. Compared with vehicles aged 16 years or more, striking vehicles under 10 years old and struck vehicles under 5 years old are associated with a higher own-driver injury probability, and striking vehicles under 10 years old are also associated with a higher struck-driver injury probability. The corresponding marginal effects range from 0.1008 to 0.1494. This result differs from severe-injury and fatality studies, which report lower risk in newer vehicles (Hoye 2019; Zhang et al. 2025). A plausible reason is that the binary outcome counts possible injuries together with serious ones, so the reporting of minor injury may outweigh differences in crashworthiness in sideswipe crashes. A similar direction has been reported before, with older vehicles associated with lower driver injury severity in urban fixed-object crashes (Yamamoto and Shankar 2004). In rear-end crashes, vehicle age was tested in the same four categories but did not improve the model fit (likelihood ratio statistic of 4.68 with nine degrees of freedom), so it is not retained.

*5.2.3. Roadway and contextual characteristics*

Among the roadway characteristics, a higher posted speed limit is associated with a higher injury probability for both drivers in rear-end crashes and for the striking driver in sideswipe crashes. This pattern is consistent with greater impact energy at higher speeds (Russo et al. 2014). For rear-end crashes, interstate segments are associated with lower injury probabilities for both drivers, while local roads and service or frontage roads are associated with higher struck-driver injury probability. Similar results have been reported for rear-end crashes, with lower injury probability on interstates in work zones (Madarshahian et al. 2025) and lower fatality risk on main roads than on parallel roads (Champahom et al.

2020). The latter study suggests that entry to the main roads is limited, so the striking and struck vehicles travel at similar speeds and the crash is less severe. On a parallel road, in contrast, the struck vehicle may be stationary, so the speed difference between the two vehicles is larger. Service or frontage roads also show the largest marginal effect among the roadway and contextual variables in rear-end crashes, a 0.1274 higher injury probability for the struck driver.

The association between intersections and injury probability differs by crash type. In sideswipe crashes, a location related to an intersection or driveway is associated with a modestly higher striking-driver injury probability, and three-leg and four-leg intersections with higher struck-driver injury probability. In rear-end crashes, however, the association between such a location and struck-driver injury probability varies across crashes, as discussed in Section 5.2.5. Signalized locations and locations with no traffic control are both associated with lower striking-driver injury probability in rear-end crashes. For signalized locations, a possible reason is that drivers slow as they approach a signal. Champahom et al. (2020) give a similar reason for the lower fatality risk of rear-end crashes at intersections. In sideswipe crashes, a roadway that is not straight is associated with lower striking-driver injury probability, perhaps because drivers travel more slowly and are more attentive on curves.

Among the contextual characteristics, darkness without lighting and adverse weather are associated with higher struck-driver injury probability in sideswipe crashes, and peak periods with lower struck-driver injury probability. Darkness without lighting also shows the largest marginal effect among the roadway and contextual variables in these crashes, a 0.1423 higher injury probability for the struck driver, followed by three-leg intersections (0.1023). In rear-end crashes, lighted darkness and weekends are associated with higher striking-driver injury probability, while the night period and peak periods are associated with lower striking-driver injury probability. For peak periods, this association holds on average but varies across crashes, as discussed in Section 5.2.5. The scene-composition measures have small marginal effects, between 0.0113 and 0.0310. Building density is associated with higher injury probability for both drivers in rear-end crashes. In sideswipe crashes, person density is associated with higher striking-driver injury probability, and terrain and road-surface density with higher struck-driver injury probability. Although these associations are small, they suggest that the visual environment of the crash location may be related to injury probability, as has been reported for crash frequency at intersections (Wang et al. 2025).

*5.2.4. ADAS functions*

The ADAS functions are examined on both the striking and the struck vehicle, giving four analytical perspectives. These are the striking driver's self-protection, the struck driver's self-protection, the protective spillover to the struck driver, and the reverse spillover to the striking driver. Figure 2 shows the marginal effects of the ADAS functions for both crash types, grouped by the three perspectives in which significant associations are found. Each marginal effect is the change in a driver's injury probability when the function is present, compared with when it is absent (the reference category of each function is given in Table 4 and Table 5).

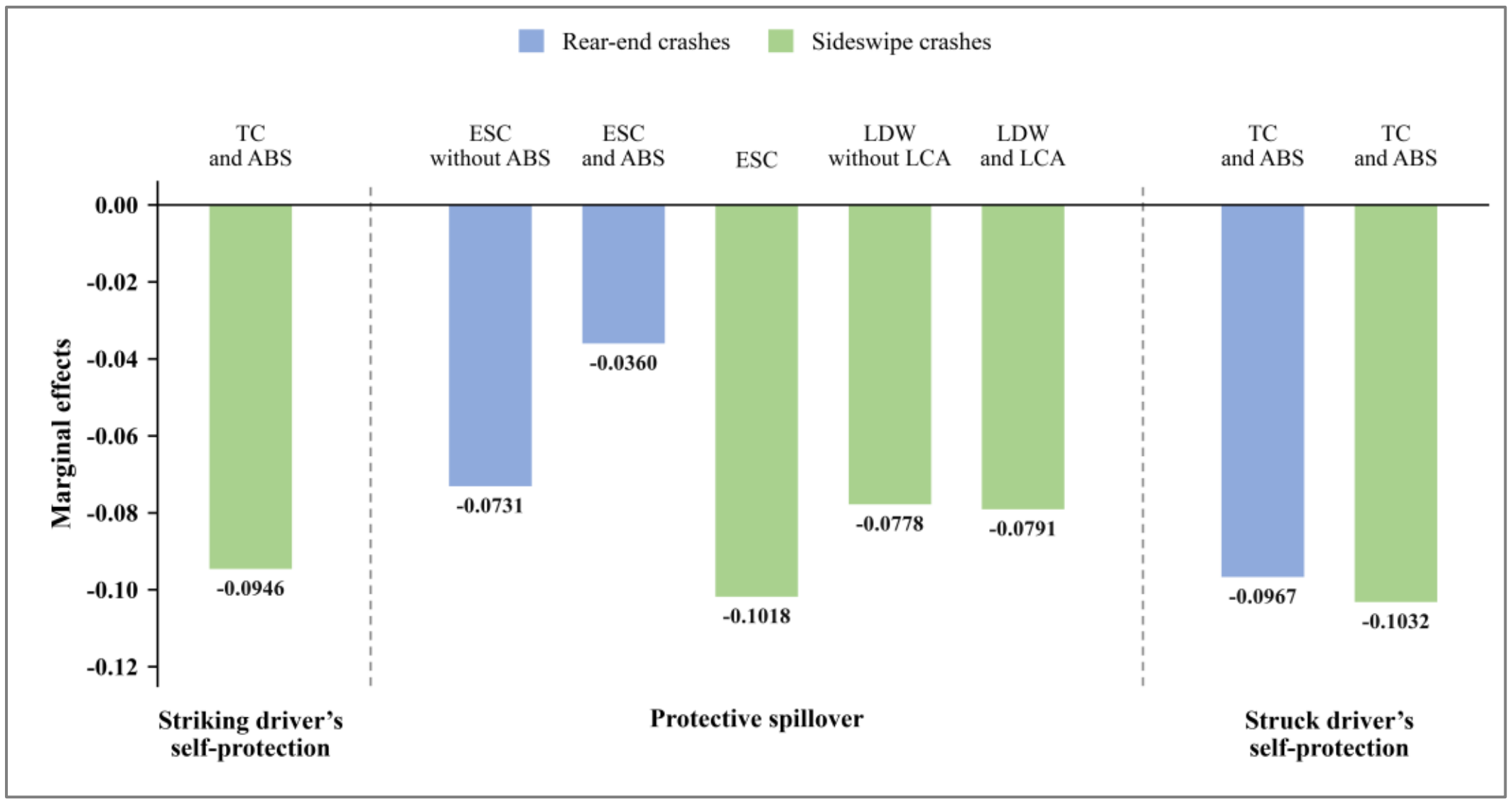

**Figure 2 Marginal effects of the ADAS functions by analytical perspective and crash type**

The protective spillover effects are among the study's most interesting findings. In rear-end crashes, ESC on the striking vehicle is associated with a lower injury probability for the struck driver, whether or not ABS is recorded with it. The marginal effects show a struck-driver injury probability lower by 0.0731 when ABS is not recorded and by 0.0360 when it is, compared with a striking vehicle without ESC. These functions help drivers avoid collisions. If a collision still occurs, the striking vehicle may have a lower speed, better alignment, and greater stability at impact, leaving the struck driver less likely to be injured (Erke 2008). ESC brakes individual wheels to keep the vehicle headed in the direction the driver intends (NHTSA 2007), and ABS keeps the wheels from locking, so the driver can still steer during hard braking and, on many surfaces, stop in a shorter distance (Sivinski 2011). A striking driver who brakes hard or steers away just before a rear-end collision can therefore stay in control and continue to slow down until the impact. Although the estimated reduction in struck-driver injury probability is larger when ABS is not recorded, this estimate is based on only 216 striking vehicles, 93% of which predate model year 2012. ESC, which incorporates ABS, became mandatory for all new light vehicles by that model year (NHTSA 2007). Given this group's small size and concentration in older vehicles, the difference between the two estimates is not interpreted further.

The same protective spillover appears in sideswipe crashes. ESC on the striking vehicle is again associated with a lower injury probability for the struck driver, and LDW has the same association, with or without LCA. The marginal effects further show that the protective spillovers are larger in sideswipe crashes than in rear-end crashes. For ESC, the struck-driver injury probability is lower by 0.1018 in sideswipe crashes, compared with a striking vehicle without ESC. A sideswipe often occurs when the striking vehicle changes lanes or drifts toward the adjacent lane, and a driver who notices the other vehicle late may steer back sharply. ESC can keep such a sharp correction from turning into a skid, so the striking vehicle is more likely to stay on its path and hit the struck vehicle at a shallow angle rather than turn into its side. Nearly all striking vehicles under 10 years old in the sideswipe sample are equipped with ESC, while about half of those aged 16 years or more are not. For the newer vehicles, the higher struck-driver injury probability associated with their age group is partly offset by the lower probability associated with ESC.

LDW on the striking vehicle shows somewhat smaller reductions than ESC in sideswipe crashes, 0.0778 without LCA and 0.0791 with it. The warning alerts the driver to an unintentional lane departure, for example one caused by distraction (Kusano et al. 2014), and LCA also steers the vehicle back toward the center of its lane. A warned driver who begins to correct the path before contact may strike the adjacent vehicle at a lower lateral speed and a shallower angle, which is consistent with the lower injury probability of the struck driver.

The struck driver's self-protection shows the largest reduction in injury probability among the ADAS functions in rear-end crashes and, by a small margin, in sideswipe crashes. TC with ABS on the struck vehicle is associated with an own-driver injury probability that is lower by 0.0967 and 0.1032, respectively. However, the sideswipe association is significant at the 90% confidence level, compared with the 99% level in rear-end crashes. The struck vehicle is pushed forward in a rear-end crash and sideways in a sideswipe, and its driver may brake or steer during or just after the impact. Because TC and ABS prevent the wheels from spinning or locking, the struck driver is more likely to stay in control instead of sliding after the impact (Sivinski 2011). In addition, these functions are more common in later-model vehicles, which may also include other safety features that improve occupant protection (Hoye 2019; Zhang et al. 2025). Because the sideswipe model controls vehicle age only in broad categories, the 0.1032 reduction may partly reflect the broader safety package of equipped vehicles, not only the named functions.

The striking driver's self-protection appears only in sideswipe crashes, and only for TC with ABS. This association corresponds to a striking-driver injury probability that is lower by 0.0946 and is significant at the 95% confidence level. Because the striking driver in a sideswipe is often steering at the moment of contact, the same functions can help keep the vehicle on its path after the contact, which may lower the striking driver's own injury probability. For the reverse spillover, the struck vehicle's functions show no statistically significant association with the striking driver's injury in either crash type. Finally, the forward-looking functions are considered only on the striking vehicle in rear-end crashes, because the struck vehicle is hit from behind. They show no significant association with either driver's injury. Their main benefit, however, is the prevention of rear-end crashes (Cicchino 2017), and prevented crashes do not appear in crash data.

*5.2.5. Random parameters*

In the rear-end model, two coefficients vary significantly across crashes, and both describe the time and location of the crash rather than the ADAS functions. The striking-driver peak-period coefficient has a mean of -0.567 and a standard deviation of 0.820, so peak periods are associated with lower injury probability for about 76% of striking drivers and higher for the rest. This spread may reflect traffic conditions that the data do not record, such as how congested the road was at the time of the crash (Mannering et al. 2016). On average, peak periods are associated with a striking-driver injury probability lower by 0.0420. The intersection-context coefficient for struck drivers has a mean close to zero (-0.039) and a significant standard deviation of 0.690, so intersection-related locations are associated with lower injury probability for about half of struck drivers and higher for the other half. Its average marginal effect, 0.0320, is positive despite the slightly negative mean. The effect is simulated over the coefficient's estimated distribution rather than evaluated at the mean, and on the nonlinear probability scale such an average need not share the mean's sign (Hou et al. 2022).

The two random parameters are treated as uncorrelated, because allowing a correlation between them does not significantly improve the fit (likelihood ratio statistic of 2.09 with one degree of freedom). Likewise, heterogeneity in the means and variances of the two parameters is not supported. The remaining variables in the rear-end model and all variables in the sideswipe model are estimated as fixed parameters. These include the ADAS functions, whose associations with injury are consistent across crashes in both crash types.

## 6. Conclusions

This study examined how the ADAS functions of the two vehicles in a same-direction crash relate to the injuries of both drivers when a collision is not avoided. Police-reported two-vehicle rear-end and same-direction sideswipe crashes in Texas from 2023 to 2025 were resolved into a striking and a struck vehicle, and the ADAS functions of each vehicle were identified at function level from vehicle identification numbers. The two drivers' injury outcomes were then estimated jointly for each crash type through a bivariate probit model that allowed random parameters where the data supported them. Each function on either vehicle could therefore be related to its own driver's injury and to the other driver's injury, capturing both self-protection and cross-vehicle spillover.

The paired design is supported by the data. In both crash types the two drivers' injuries are significantly correlated through unobserved crash factors, which justifies modeling them jointly, with random parameters retained for rear-end crashes and fixed parameters for sideswipe crashes. For the rear-end model, the richer random parameters structures, such as correlated parameters and heterogeneity in the means or variances, were examined and none was supported. The random parameters apply only to contextual factors, not to the ADAS functions, so the ADAS associations are consistent across crashes. A single model per crash type over the full study period is likewise supported, with a likelihood ratio test finding no evidence against pooling the study years.

The self-protection and cross-vehicle spillover perspectives reveal an asymmetric set of ADAS function associations. The protective spillover effect is the study's central pattern. ESC on the striking vehicle is associated with a lower injury probability for the struck driver in both crash types. LDW on the striking vehicle shows the same protective spillover effect in sideswipe crashes, with or without LCA. TC with ABS on the struck vehicle is associated with a lower injury probability for its own driver in both crash types, the largest reduction among the ADAS functions in each. In sideswipe crashes, this association may also reflect other occupant-protection features of the vehicles that have these functions. On the striking vehicle, the same functions are associated with a lower injury probability for its own driver only in sideswipe crashes. No reverse spillover appears, since the struck vehicle's functions show no significant association with the striking driver's injury in either crash type. Among the other characteristics, vehicle size follows the familiar compatibility pattern, with a larger vehicle associated with a lower injury probability for its own driver in both crash types and, in rear-end crashes, also with a higher probability for the other driver. Driver, roadway, and contextual characteristics such as impairment, posted speed, and darkness are also significantly associated with injury probability.

Protective associations differ across the ADAS function families. Lower injury probabilities are associated with the stability functions and, in sideswipe crashes, the lateral warning functions. The forward-looking functions mainly prevent rear-end crashes, and prevented crashes are not in the sample. The stability and lateral warning functions, by contrast, can still reduce the speed or improve the alignment at impact in a crash they do not prevent, which may lower the resulting injury.

The study thus identifies indirect associations between the two vehicles' function-level ADAS availability and the injuries of both drivers across the two crash types. These associations are conditional on a crash having occurred and should not be interpreted as causal effects. Based on these findings, several practical implications can be drawn. The protective spillover effects reach the struck driver, so the lower injury probabilities associated with stability and lateral warning functions appear to extend beyond the vehicles equipped with them. A fleet in which these functions are more widespread may therefore see lower injury probabilities among drivers who do not themselves have them. Evaluations of these functions by safety agencies and vehicle manufacturers could also consider the injuries of the other driver, since an assessment limited to the occupants of the equipped vehicle would not capture this association. In addition, the roadway and contextual results may help transportation agencies in developing countermeasures for rear-end and sideswipe crashes.

This study should be viewed with a few points in mind. Although function coding identifies standard equipment rather than activation in a specific crash, Cox et al. (2025) found that automatic emergency braking and lane departure systems were active in 93% and 87% of observed equipped vehicles, respectively. The present analysis focuses on one state, two-vehicle same-direction crashes from 2023 to 2025, and a binary injury outcome. Future work can extend this framework to other states and crash configurations, examine finer injury-severity levels where the data permit, and assess temporal stability over longer study periods.

**CRediT authorship contribution statement**

**Parvez Anowar:** Writing – original draft, Writing – review & editing, Methodology, Formal analysis, Data curation, Conceptualization. **Mohamed Abdel-Aty:** Writing – review & editing, Writing – original draft, Supervision, Methodology, Formal analysis, Conceptualization. **Chenzhu Wang:** Writing – review & editing, Writing – original draft, Methodology, Formal analysis, Conceptualization. **Shaoyan Zhai:** Data curation.

**Declaration of competing interest**
The authors have no competing interests to declare.

**Funding**
This research did not receive any specific grant from funding agencies in the public, commercial, or not-for-profit sectors.

**Data availability**
The data used in this study are available from the corresponding author on request.

**Appendix A**

**Table A1 Descriptive statistics of the model variables for rear-end and sideswipe crashes**

| Variable | Rear-end (N = 2,273) | | | | Sideswipe (N = 1,216) | | | |
|---|---|---|---|---|---|---|---|---|
| | Mean | SD | Min | Max | Mean | SD | Min | Max |
| **Driver characteristics** | | | | | | | | |
| V1 male driver | 0.601 | 0.490 | 0 | 1 | 0.546 | 0.498 | 0 | 1 |
| V2 male driver | 0.476 | 0.500 | 0 | 1 | 0.533 | 0.499 | 0 | 1 |
| V1 driver < 25 | 0.304 | 0.460 | 0 | 1 | 0.249 | 0.433 | 0 | 1 |
| V1 driver 25 to 64 | 0.626 | 0.484 | 0 | 1 | 0.636 | 0.481 | 0 | 1 |
| V1 driver ≥ 65 | 0.069 | 0.254 | 0 | 1 | 0.115 | 0.319 | 0 | 1 |
| V2 driver < 25 | 0.151 | 0.358 | 0 | 1 | 0.162 | 0.369 | 0 | 1 |
| V2 driver 25 to 64 | 0.764 | 0.425 | 0 | 1 | 0.751 | 0.433 | 0 | 1 |
| V2 driver ≥ 65 | 0.085 | 0.279 | 0 | 1 | 0.087 | 0.282 | 0 | 1 |
| V1 driver unlicensed | 0.121 | 0.326 | 0 | 1 | 0.115 | 0.319 | 0 | 1 |
| V2 driver unlicensed | 0.064 | 0.244 | 0 | 1 | 0.065 | 0.247 | 0 | 1 |
| V1 driver unrestrained | 0.007 | 0.086 | 0 | 1 | 0.002 | 0.041 | 0 | 1 |
| V2 driver unrestrained | 0.002 | 0.047 | 0 | 1 | 0.002 | 0.041 | 0 | 1 |
| V1 driver impaired | 0.013 | 0.112 | 0 | 1 | 0.005 | 0.070 | 0 | 1 |
| V2 driver impaired | 0.001 | 0.030 | 0 | 1 | 0.000 | 0.000 | 0 | 0 |
| V1 one occupant | 0.782 | 0.413 | 0 | 1 | 0.732 | 0.443 | 0 | 1 |
| V1 two occupants | 0.143 | 0.350 | 0 | 1 | 0.187 | 0.390 | 0 | 1 |
| V1 three or more occupants | 0.075 | 0.264 | 0 | 1 | 0.081 | 0.274 | 0 | 1 |
| V2 one occupant | 0.675 | 0.468 | 0 | 1 | 0.697 | 0.460 | 0 | 1 |
| V2 two occupants | 0.213 | 0.410 | 0 | 1 | 0.202 | 0.402 | 0 | 1 |
| V2 three or more occupants | 0.111 | 0.315 | 0 | 1 | 0.100 | 0.301 | 0 | 1 |
| **Vehicle characteristics** | | | | | | | | |
| V1 larger body | 0.548 | 0.498 | 0 | 1 | 0.555 | 0.497 | 0 | 1 |
| V2 larger body | 0.580 | 0.494 | 0 | 1 | 0.558 | 0.497 | 0 | 1 |
| V1 vehicle age under 5 years | 0.276 | 0.447 | 0 | 1 | 0.414 | 0.493 | 0 | 1 |
| V1 vehicle age 5 to 9 years | 0.295 | 0.456 | 0 | 1 | 0.273 | 0.446 | 0 | 1 |
| V1 vehicle age 10 to 15 years | 0.229 | 0.420 | 0 | 1 | 0.178 | 0.383 | 0 | 1 |
| V1 vehicle age 16 years or more | 0.200 | 0.400 | 0 | 1 | 0.135 | 0.342 | 0 | 1 |
| V2 vehicle age under 5 years | 0.580 | 0.494 | 0 | 1 | 0.516 | 0.500 | 0 | 1 |
| V2 vehicle age 5 to 9 years | 0.240 | 0.427 | 0 | 1 | 0.285 | 0.452 | 0 | 1 |
| V2 vehicle age 10 to 15 years | 0.113 | 0.316 | 0 | 1 | 0.115 | 0.319 | 0 | 1 |
| V2 vehicle age 16 years or more | 0.068 | 0.251 | 0 | 1 | 0.083 | 0.276 | 0 | 1 |

| Variable | Rear-end (N = 2,273) | | | | Sideswipe (N = 1,216) | | | |
|---|---|---|---|---|---|---|---|---|
| | Mean | SD | Min | Max | Mean | SD | Min | Max |
| **Roadway and contextual characteristics** | | | | | | | | |
| Posted speed limit (mph) | 49.0 | 12.9 | 10.0 | 80.0 | 47.1 | 13.0 | 10.0 | 85.0 |
| Intersection context | 0.455 | 0.498 | 0 | 1 | 0.385 | 0.487 | 0 | 1 |
| Interstate roadway | 0.264 | 0.441 | 0 | 1 | 0.276 | 0.447 | 0 | 1 |
| US or state highway | 0.433 | 0.496 | 0 | 1 | 0.346 | 0.476 | 0 | 1 |
| Local roadway | 0.303 | 0.460 | 0 | 1 | 0.377 | 0.485 | 0 | 1 |
| Main lane | 0.883 | 0.322 | 0 | 1 | 0.822 | 0.382 | 0 | 1 |
| Service road | 0.085 | 0.279 | 0 | 1 | 0.153 | 0.360 | 0 | 1 |
| Ramp, connector, or other | 0.032 | 0.175 | 0 | 1 | 0.025 | 0.155 | 0 | 1 |
| Traffic control signalized | 0.236 | 0.425 | 0 | 1 | 0.120 | 0.325 | 0 | 1 |
| Traffic control lane markings | 0.622 | 0.485 | 0 | 1 | 0.740 | 0.439 | 0 | 1 |
| Traffic control stop, yield, or warning sign | 0.071 | 0.257 | 0 | 1 | 0.057 | 0.231 | 0 | 1 |
| Traffic control none or uncontrolled | 0.055 | 0.229 | 0 | 1 | 0.066 | 0.248 | 0 | 1 |
| Traffic control other | 0.015 | 0.123 | 0 | 1 | 0.017 | 0.130 | 0 | 1 |
| Three-leg intersection | 0.083 | 0.276 | 0 | 1 | 0.101 | 0.302 | 0 | 1 |
| Four-leg intersection | 0.097 | 0.296 | 0 | 1 | 0.137 | 0.343 | 0 | 1 |
| Other or not an intersection | 0.820 | 0.385 | 0 | 1 | 0.762 | 0.426 | 0 | 1 |
| Roadway not straight | 0.047 | 0.212 | 0 | 1 | 0.070 | 0.255 | 0 | 1 |
| Adverse weather | 0.233 | 0.423 | 0 | 1 | 0.223 | 0.416 | 0 | 1 |
| Daylight | 0.834 | 0.372 | 0 | 1 | 0.831 | 0.375 | 0 | 1 |
| Dawn, dusk, or lighting unknown | 0.030 | 0.172 | 0 | 1 | 0.027 | 0.163 | 0 | 1 |
| Dark, lighted | 0.099 | 0.299 | 0 | 1 | 0.109 | 0.312 | 0 | 1 |
| Dark, not lighted | 0.037 | 0.188 | 0 | 1 | 0.032 | 0.176 | 0 | 1 |
| Crash time peak | 0.484 | 0.500 | 0 | 1 | 0.439 | 0.496 | 0 | 1 |
| Crash time midday | 0.392 | 0.488 | 0 | 1 | 0.423 | 0.494 | 0 | 1 |
| Crash time night | 0.125 | 0.330 | 0 | 1 | 0.138 | 0.345 | 0 | 1 |
| Weekend | 0.179 | 0.383 | 0 | 1 | 0.178 | 0.383 | 0 | 1 |
| **Scene composition (percent of scene pixels)** | | | | | | | | |
| Building density | 6.2 | 10.3 | 0.0 | 98.8 | 6.1 | 9.7 | 0.0 | 75.3 |
| Terrain | 6.5 | 6.0 | 0.0 | 40.6 | 6.5 | 6.2 | 0.0 | 32.7 |
| Road surface density | 33.5 | 9.2 | 0.0 | 55.0 | 33.0 | 9.5 | 0.2 | 52.4 |
| Person density | 0.1 | 1.1 | 0.0 | 31.0 | 0.1 | 0.2 | 0.0 | 4.0 |
| **ADAS functions** | | | | | | | | |
| V1 ESC without ABS | 0.095 | 0.293 | 0 | 1 | 0.118 | 0.322 | 0 | 1 |
| V1 ESC and ABS | 0.714 | 0.452 | 0 | 1 | 0.803 | 0.398 | 0 | 1 |
| V1 ESC | 0.809 | 0.393 | 0 | 1 | 0.921 | 0.270 | 0 | 1 |
| V1 TC and ABS | 0.711 | 0.453 | 0 | 1 | 0.803 | 0.398 | 0 | 1 |
| V1 LDW without LCA | 0.220 | 0.414 | 0 | 1 | 0.447 | 0.497 | 0 | 1 |
| V1 LDW and LCA | 0.088 | 0.284 | 0 | 1 | 0.169 | 0.375 | 0 | 1 |
| V2 TC and ABS | 0.893 | 0.310 | 0 | 1 | 0.873 | 0.333 | 0 | 1 |

Note: V1 = striking-vehicle, and V2 = struck-vehicle.

**Table A2 Rear-end model estimates under alternative distributions of the random parameters**

| Distribution | Model statistics | | | | | Model estimates (t-stat) | | |
|---|---|---|---|---|---|---|---|---|
| | Log-likelihood | AIC | BIC | Cross-equation correlation | Vuong z | V1 ESC without ABS | V1 ESC and ABS | V2 TC and ABS |
| Normal | -1871.67 | 3825.3 | 4060.2 | 0.559 | - | -0.317 (-2.405) | -0.139 (-1.685) | -0.348 (-3.377) |
| Triangular | -1871.65 | 3825.3 | 4060.2 | 0.560 | -0.19 | -0.317 (-2.405) | -0.139 (-1.686) | -0.349 (-3.382) |

| Distribution | Model statistics | | | | | Model estimates (t-stat) | | |
|---|---|---|---|---|---|---|---|---|
| | Log-likelihood | AIC | BIC | Cross-equation correlation | Vuong z | V1 ESC without ABS | V1 ESC and ABS | V2 TC and ABS |
| Uniform | -1871.61 | 3825.2 | 4060.1 | 0.562 | -0.21 | -0.317 (-2.397) | -0.139 (-1.682) | -0.351 (-3.391) |

Note: All models have 41 parameters and are estimated with 1,000 Halton draws. The Vuong z compares each distribution with the normal distribution. V1 = striking-vehicle, and V2 = struck-vehicle.